\documentclass[traditabstract]{aa}  
\usepackage{graphicx}
\usepackage{txfonts}
\usepackage{rotating}
\usepackage{natbib}
\usepackage[usenames]{color}
\usepackage{multirow}
\usepackage{url}
\usepackage{ulem}
\bibpunct{(}{)}{;}{a}{}{,} 

\newcommand{\kms}   {{\rm \  km \  s^{-1}}}
\newcommand{\mspc}   {{\rm \  m \  s^{-1} \  pc^{-1}}}

\newcommand{\pc}   {{\rm \  pc}}

\newcommand{\msun}{M$_{\odot}$}   
\newcommand{\msol}{{\rm M}_{\odot}}   

\begin{document}

\title{ Rotation of molecular clouds in M~51 }
		
\author{J. Braine\inst{1} \and A. Hughes\inst{2}  \and E. Rosolowsky\inst{3} \and P. Gratier\inst{1} \and D. Colombo{\inst4} \and S. Meidt\inst{4} \and  E. Schinnerer\inst{5} }

\institute{Laboratoire d'Astrophysique de Bordeaux, Univ. Bordeaux, CNRS, B18N, all\'ee Geoffroy Saint-Hilaire, 33615 Pessac, France.\\
             \email{jonathan.braine@u-bordeaux.fr}
        \and Institut de Recherche en Astrophysique et Plan\'etologie (IRAP), Universit\'e Paul Sabatier, Toulouse cedex 4, France
        \and Department of Physics, University of Alberta, 4-183 CCIS, Edmonton, AB, T6G 2E1, Canada
        \and Max-Planck-Institut f\"ur Radioastronomie, Auf dem H\"ugel 69, 53121 Bonn, Germany
      \and  Max-Planck-Institut for Astronomy, Konigstuhl 17, 69117 Heidelberg, Germany  }
\date{Received Nov 9, 2018; accepted Nov. 18, 2019}

\abstract { The grand-design spiral galaxy M~51 was observed at 40pc resolution in CO(1--0) by the PAWS project. 
A large number of molecular clouds were identified and we search for velocity gradients in two high signal-to-noise subsamples, containing 682 and 376 clouds.  
The velocity gradients are found to be systematically prograde oriented, as was previously found for the rather flocculent spiral M~33.  This strongly supports the idea that the velocity gradients reflect cloud rotation, rather than more random dynamical forces, such as turbulence.
Not only are the gradients prograde, but their $\frac{\partial v}{\partial x}$ and $\frac{\partial v}{\partial y}$ coefficients follow galactic shear in sign, although with a lower amplitude.  No link is found between the orientation of the gradient and the orientation of the cloud.  The values of the cloud angular momenta appear to be an extension of the values noted for galactic clouds despite the orders of magnitude difference in cloud mass.  
Roughly 30\% of the clouds show retrograde velocity gradients.  For a strictly rising rotation curve, as in M~51, gravitational contraction would be expected to yield strictly prograde rotators within an axisymmetric potential. In M~51, the fraction of retrograde rotators is found to be higher in the spiral arms than in the disk as a whole.  Along the leading edge of the spiral arms, a majority of the clouds are retrograde rotators.
While this work should be continued on other nearby galaxies, the M~33 and M~51 studies have shown that clouds rotate and that they rotate mostly prograde, although the amplitudes are not such that rotational energy is a significant support mechanism against gravitation.  In this work, we show that retrograde rotation is linked to the presence of a spiral gravitational potential.  
  }	


\maketitle

\section{Introduction}

Assuming giant molecular clouds (GMCs) form from the gravitational contraction of a more diffuse medium, their angular momenta should reflect the dynamics (angular momentum in particular) of the medium from which they formed. Despite the importance of measuring cloud rotation (thus angular momentum) to understand the formation mechanism of GMCs, relatively little work has appeared. The initial work was done by \citet{Kutner77}, followed by \citet{Blitz93} and works cited therein, all for molecular clouds in the Milky Way.  These authors used velocity gradients as a measure of cloud rotation.  From these studies, it was difficult to determine whether molecular clouds rotated in a systematic way or not.  \citet{Bally89} argued that at least some velocity gradients may be due to pressure from stellar winds generated by massive star formation.  \citet{Burkert2000} found that turbulence could also produce velocity gradients similar to those observed.  Most Galactic clouds have quite weak velocity gradients \citep{Blitz93}.  In their compilation, \citet{Phillips99} found that clouds tend to have angular momentum vectors oriented perpendicular to the plane of the Galaxy.  However, the number of clouds was quite small (48 in their Fig. 1a).  Any systematic effects would tend to support the idea that velocity gradients generally trace rotation rather than more random effects, such as cloud-star interactions.  Neither \citet{Blitz93} nor \citet{Phillips99} identifed a pro- or retrograde orientation of the velocity gradients observed in the 
Galaxy.  "Pro-" or "retro"-grade rotation is defined with respect to the sense of rotation of the galaxy (prograde being in the same direction as galactic rotation), rather than a local gravitational potential.

It is now possible to observe molecular clouds in nearby galaxies.  The first to look at cloud rotation in nearby galaxies were \citet{Rosolowsky03}, followed by \citet{Imara11b}, from their observations of M~33.  The velocity gradients were weak in their sample of 45 clouds.  The deep CO(2--1) survey with the Institut de Radioastronomie Millim\'etrique (IRAM) 30meter telescope \citep{Druard14, Gratier10} and the identification of a large sample of molecular clouds \citep{Corbelli17,Gratier12} enabled \citet{Braine2018}, hereafter B18, to measure velocity gradients.  B18 show that the velocity gradients of the clouds are systematically in the prograde direction, lending support to the idea that the gradients are due to rotation.  The rotational energy is, however, very low and not a significant source of support.

\begin{figure}
	\centering
	\includegraphics[width=\hsize{}]{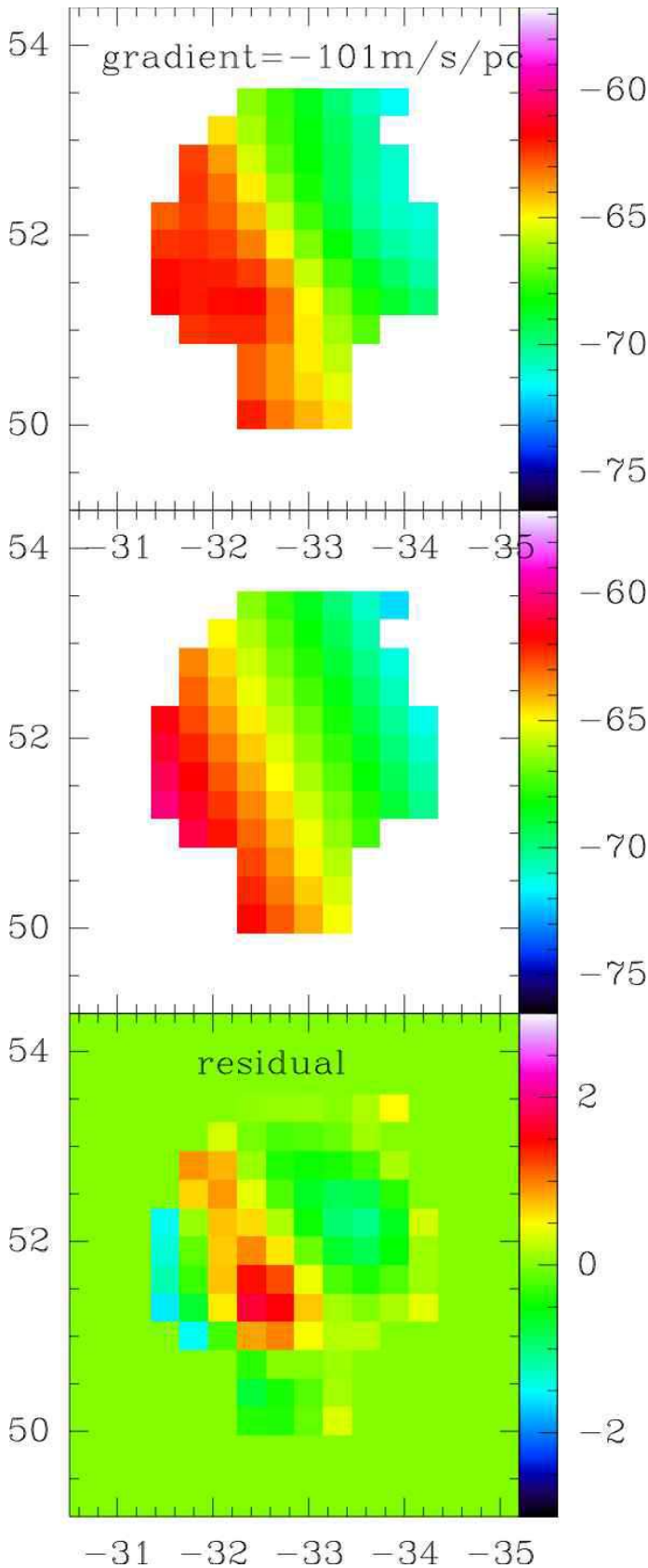}
	\caption{First moment velocities (top panel) followed by the calculated gradient and the residual image.  The color wedges show the velocities in $\kms$ and the $x$ and $y$ axes show the position of the cloud in arcseconds. }
	\label{cloud1417} 
\end{figure}

\begin{figure}
	\centering
	\includegraphics[width=\hsize{}]{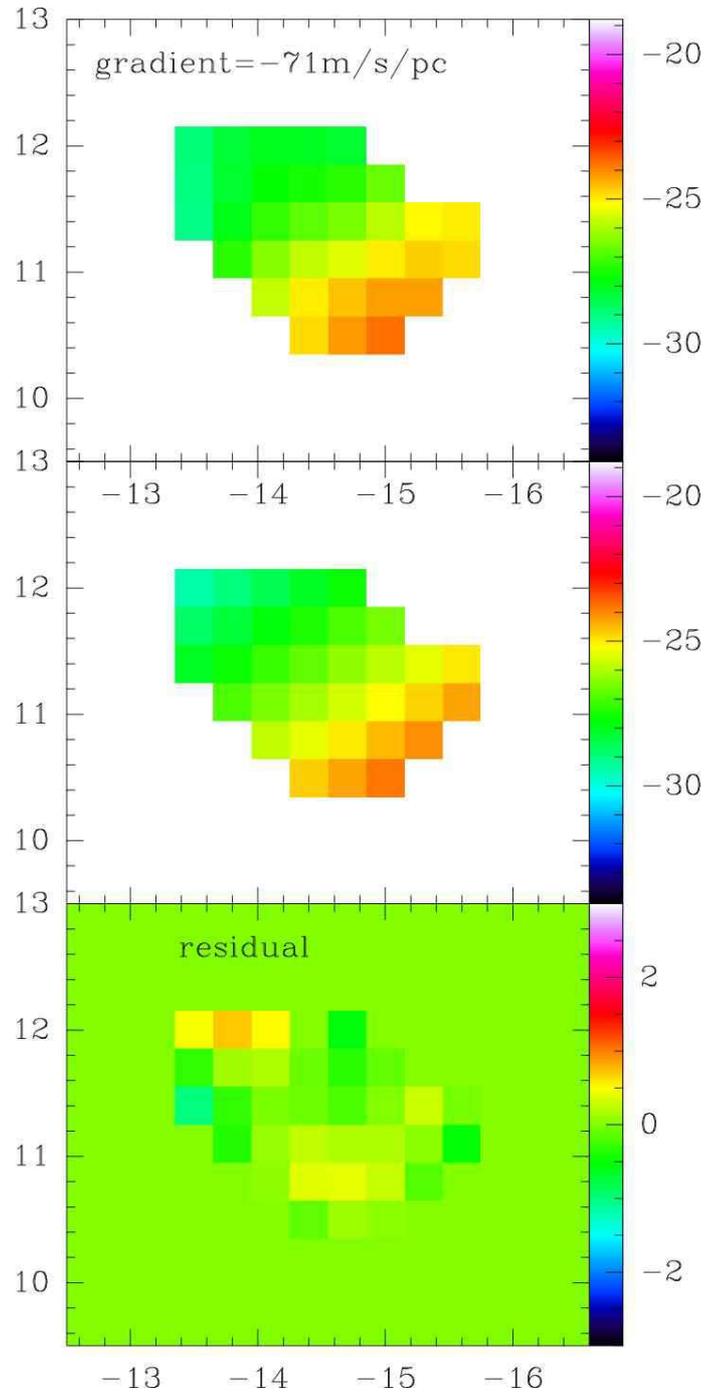}
	\caption{Same as Fig. \ref{cloud1417} but for a smaller cloud with a smaller gradient. }
	\label{cloud1005} 
\end{figure}

Here we use the high-resolution CO(1--0) PAWS observations of M~51 \citep[PdBI Arcsecond Whirlpool Survey,][]{Schinnerer2013} and the cloud identifications from \citet{Colombo2014a} to explore cloud rotation.  The datacubes and images are available from the MPIA and IRAM websites \footnote{{\tt http://www.iram.fr/ILPA/LP003/}} and the cloud catalog in \citet{Colombo2014a} and via ApJ.  The published and publicly available cloud catalog provides (among other quantities) cloud sizes, linewidths, and an indicative signal-to-noise (S/N) ratio.  The latter provides a good indication of the quality of the gradient calculation and is used to define our two cloud samples.
M~51 is often thought of as a face-on spiral but the observed rotation curve reaches about $100\kms$ so in-plane rotation is visible.  The inclination is about $22^\circ$ \citep{Colombo2014b}, but with significant uncertainty \citep{Hu2013}.  
Despite the velocity resolution of $5 \kms$, the clouds are detected in several adjacent channels.  The situation is similar to M~33 where the velocity resolution was $2.6 \kms$, but cloud linewidths in M~51 are roughly twice as broad for an equivalent cloud size (see B18 Fig. 3 for a comparison).

Coincidentally similar to M~33, the major axis is oriented close to N-S and the northern side is approaching, such that the velocity gradient is positive toward the south.  The observed velocities are also similar, reaching approximately $\pm 90\kms$ over the regions observed.


\citet{Colombo2014a} used CPROPS \citep{Rosolowsky06} to identify 1507 clouds. Each cloud has an estimated S/N ratio and they define a high-quality cloud sample to have few or no false identifications.
Because high S/N is critical to detecting rotation, we define two restrictive samples within their samples.  The samples contain 682 clouds with S/N$>7$ and 376 clouds with S/N$>10$, respectively.  The value of S/N is given in column 6 of Table 1 of \citet{Colombo2014a}.   

\begin{figure}
	\centering
	\includegraphics[width=\hsize{}]{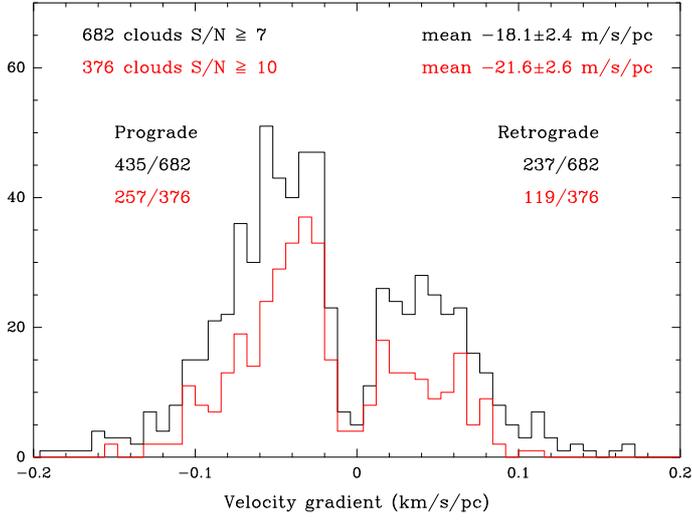}
	\caption{Histogram of velocity gradients measured for M~51 clouds.  The black histogram is for the subsample S/N $\ge 7$ and the red for the subsample S/N $\ge 10$.  The gradients in the higher S/N subsample are better defined (more prograde and with a more significant gradient). }
	\label{velgrad} 
\end{figure}

\begin{figure}
	\centering
	\includegraphics[width=\hsize{}]{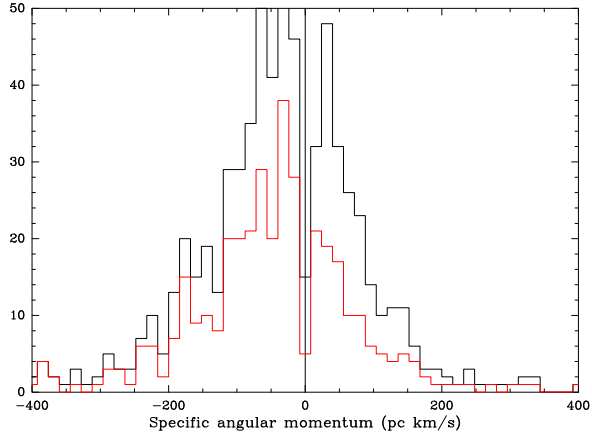}
	\caption{Histogram of angular momenta for M~51 clouds.  In black for the subsample S/N $\ge 7$ and in red for the subsample S/N $\ge 10$.  Note that the gradients in the higher S/N subsample are better defined (more prograde and with a more significant gradient).  We have included a correction for beam-smearing as in B18.  The angular momenta of the M~51 clouds are substantially higher than in M~33 (see middle panel of next figure).}
	\label{angmom} 
\end{figure}

\begin{figure}
	\centering
	\includegraphics[width=\hsize{}]{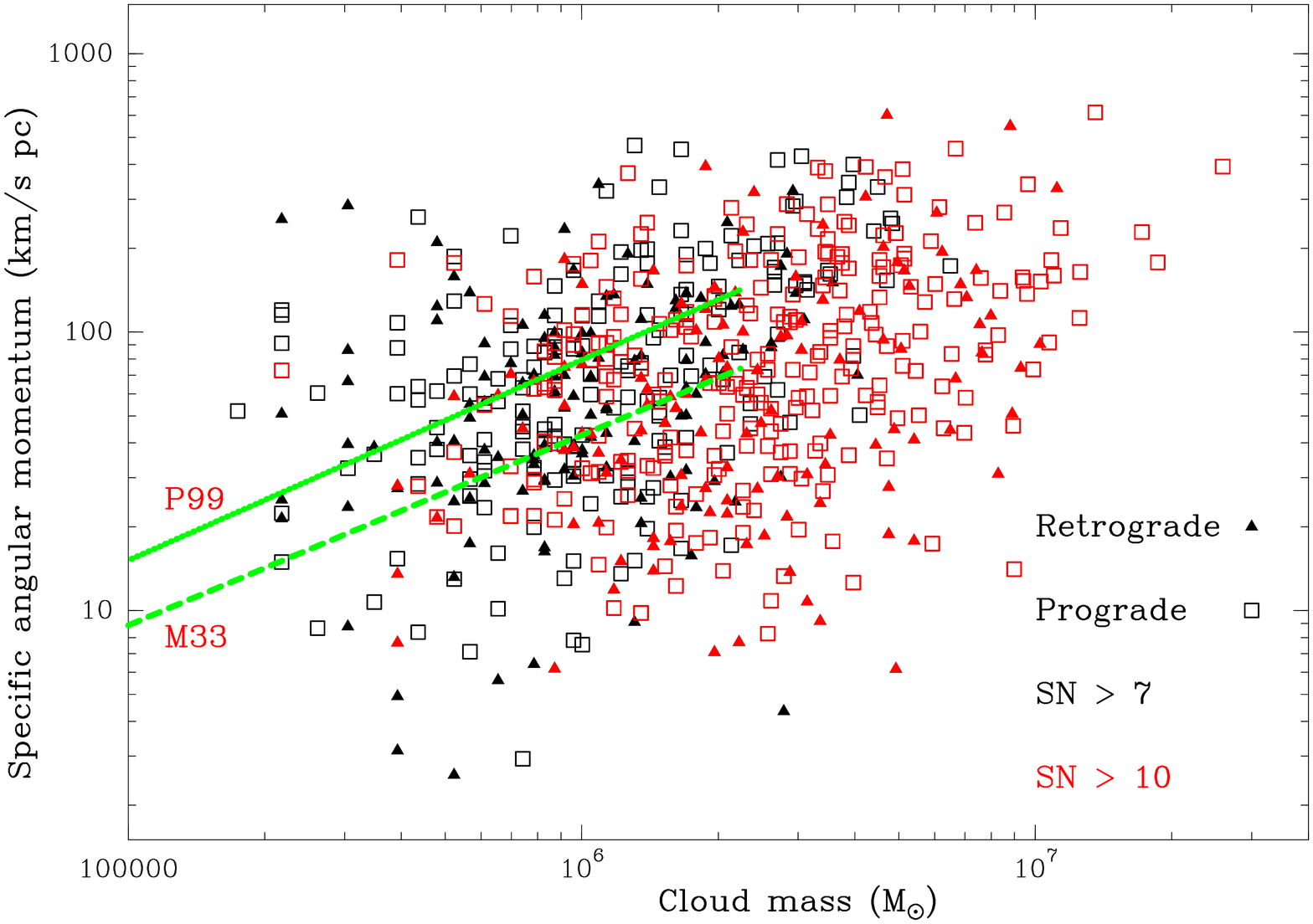}
	\includegraphics[width=\hsize{}]{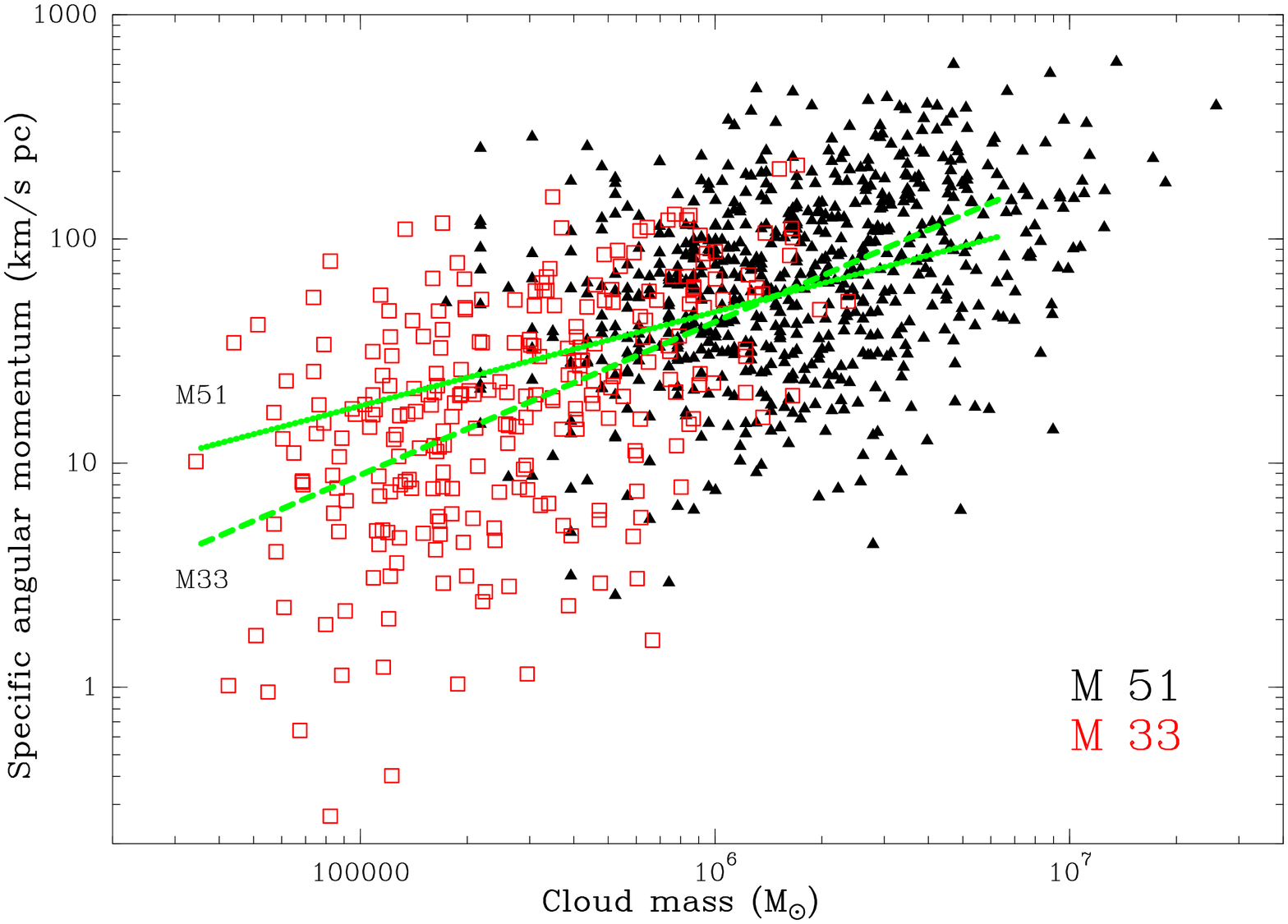}
	\includegraphics[width=\hsize{}]{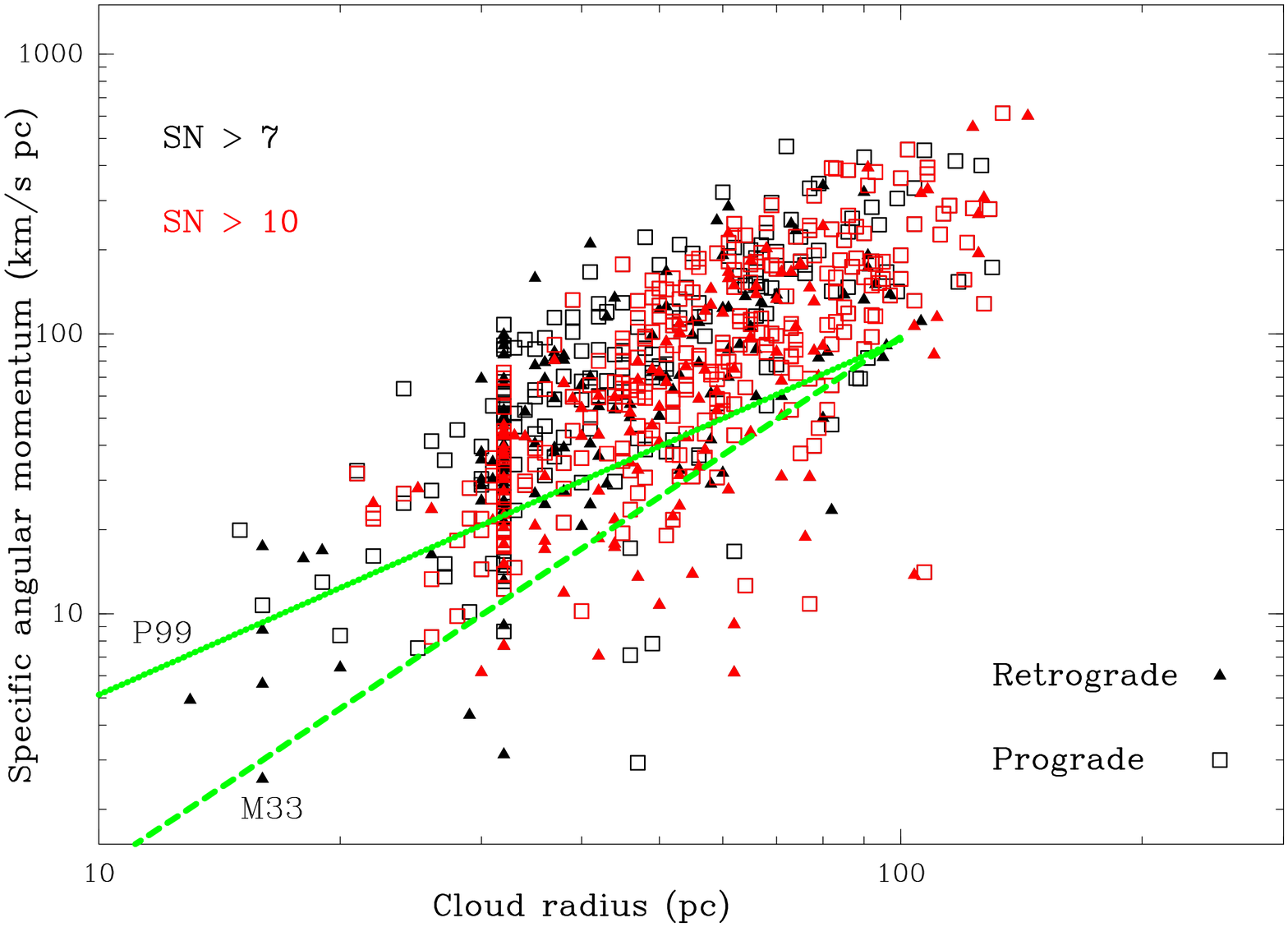}
	\caption{(top) Specific angular momentum as a function of cloud mass.  The trends from B18 for M33 and from \citet{Phillips99} are indicated.  This figure can be directly compared with B18 and with \citet{Dobbs08}.
	(middle) Here we show the strong M33 clouds along with the M~51 clouds, without showing the distinctions between pro/retrograde and S/N level.  While the values are lower for the M~33 clouds, the M~51 clouds follow a similar trend.
	(bottom) Specific angular momentum as a function of cloud radius.  The trend from \citet{Phillips99} is indicated.  The cloud radii are from \citet{Colombo2014a} and the accumulation at 32pc is due to a limit set by \citet{Colombo2014a}.  The fit results for M~33 and M~51 for both the high S/N and full samples are given in Table 1.
	The scatter about the fits is a factor 2--3 (precise values given in Table 1).}
	\label{angmom2} 
\end{figure}

\section{Velocity gradients and method to obtain them}

We follow the method used by previous authors, and described in B18 Figure 1, Figures \ref{cloud1417} and \ref{cloud1005}, and Section 3.
Each cloud is defined as a set of pixels in position-position-velocity space (i.e., a datacube) and the cloud velocity centroid is determined by CPROPS.  For each of the spatial pixels occupied by the cloud, we calculate the first moment velocity
using
\begin{equation}
v_{(x,y)} = \sum_{i=cen-2}^{i=cen+2} v_i T_i dv / \sum_{i=cen-2}^{i=cen+2} T_i dv  
\end{equation}
where $cen$ represents the central channel number.  Five velocity channels are used to calculate the velocity; the channel corresponding to the cloud velocity calculated by CPROPS and two channels to either side so that a total of $25\kms$ is covered.  The lines of the M~51 clouds are quite broad, significantly broader than those of M~33 (see Fig 3 of B18), so five channels are necessary.  Including velocities further from that of the cloud simply adds noise to the calculation.  
Once the velocity for each pixel of a cloud is known, we fit a plane to the velocities, such that the velocity gradients 
$\frac{\partial v}{\partial x}=-\frac{\partial v}{\partial RA}$ and $\frac{\partial v}{\partial y}=\frac{\partial v}{\partial Dec}$ are determined.
Figure \ref{cloud1417} illustrates the procedure.  The top panel shows the result of the first moment calculation (i.e., the representative velocity of each pixel), the middle panel shows the gradient deduced by our procedure, and the bottom panel shows the residuals.  Cloud 1417 is a large cloud with a large gradient and, unlike smaller clouds, there is structure to be seen in the residual.  The pattern of the residual image shows that rigid rotation is not appropriate as the true gradient is steeper near the cloud center and flattens somewhat.  Figure \ref{cloud1005} shows a smaller cloud with a smaller gradient.  In both cases, the values of the residual image, which can be taken as indicative of the uncertainties, are far below the total velocity gradient.

Given the $\sim 40$pc and $5\kms$ resolution of the observations, we have not attempted to fit non-linear velocity gradients.  The velocity gradients are presented in Fig. \ref{velgrad}, showing the two samples separately.  As for M~33, the prograde orientation of the high S/N sample is stronger, confirming the prograde predominance.  Due to the orientation of M~51, with a velocity increasing toward the south, prograde rotation is shown with a negative sign.  
The prograde orientation is highly significant ($8\sigma$ for the 682 cloud sample and $9\sigma$ for the 376 S/N$>10$ clouds).

Only a fraction of the velocity width can be explained via the observed velocity gradient.  $0.05 \kms$pc$^{-1}$, a reasonable value from Fig. \ref{velgrad}, over a diameter of 60pc, only corresponds to $3\kms$ whereas the typical velocity width for clouds of that size in M~51 is about $\Delta V_{fwhm} = 14\kms$.  Some of the clouds are quite large; 26 of the 376 clouds of the S/N$>10$ sample have radii of 100pc or slightly greater.  When inspecting morphologies, clearly some of these could potentially be agglomerations of smaller clouds, whether real agglomerations or simply line-of-sight associations, and approximately 1/4 are filamentary (axis ratios are given in \citet{Colombo2014a}).  

\subsection{Rotational versus binding energy}

Since velocity gradients expressed as $\kms$pc$^{-1}$ are akin to an angular frequency, we can calculate 
rotation periods and compare rotational and gravitational potential energy.
From Fig. \ref{velgrad}, we see that the vast majority of the gradients are below $0.1\kms$pc$^{-1}$ 
in absolute value although we only measure a projected gradient.  The rotational energy  
is $E=\alpha M R^2 \Omega^2$ where $\alpha$ takes values of 1/4 for a homogeneous disk, $1/6$ for a disk whose surface density declines as $r^{-1}$, 1/5 for a homogeneous sphere and less for more centrally concentrated spheres.  
Adopting $R=45$pc and $M=10^6 $\msun\ as typical values, we obtain rotational energies of $6 -10 \times 10^{48}$ ergs and gravitational binding energies of $1.1-1.9 \times 10^{51}$ ergs ($U=\beta M^2 /R$ with $\beta \sim 0.6 - 1$.  It is unlikely that the gradients are underestimated by a factor 4 or more 
so the rotational energy is well below the binding energy.  This is particularly true for centrally concentrated clouds.  

\citet{Phillips99} proposed the ratio of the rotational and turbulent virial terms as a measure of the importance of rotation:
$\beta = 0.739 \, R_{pc}^2 \, \Delta V_{fwhm}^{-2} \, \Omega^2 $ (his Eq. 1 converted to the units used here).
For representative values of $R=45$pc, $\Omega = 0.1 \kms$pc$^{-1}$ (after correction for the beam-smearing below), 
and $\Delta V_{fwhm} = 14\kms$ (slightly below average for M~51 clouds -- see Fig. 3 of B18), we obtain $\beta \approx 0.076$.  As also found by \citet{Phillips99}, this is considerably below unity, even allowing for projection effects.

If $\Omega = 0.1\kms$pc$^{-1}$ is an appropriate gradient, then the rotation period is $T=2 \pi / \Omega = 61$ Myr.
Even allowing for projection effects and the correction below, the rotation period is greater than typical GMC lifetimes of $\sim 15$Myr \citep[e.g.,][]{Corbelli17}.

\subsection{Angular momenta}

Figure \ref{angmom} shows the specific angular momenta of the clouds as a histogram. 
To calculate the angular momentum, we have assumed that the clouds can be represented as a disk with a linearly decreasing surface density, such that $j=(\Omega / 0.59) \, R^2 \, / 3$ where the factor 0.59 corrects for the underestimate in the velocity gradient due to beam smearing (see B18).  \citet{Blitz93} takes $j=\Omega R^2 / 2$, which is appropriate for a constant surface density.  Neither approximation is ideal but these expressions give an idea of the uncertainties. 

The M~51 clouds have specific angular momenta three times those found in M~33 (Fig.  \ref{angmom} and top of Fig. 12 in B18).  However, the angular momenta are similar for a given cloud mass (Fig.  \ref{angmom2}) and there is a large dispersion in all cases.  The M51 clouds are more massive, with both higher CO luminosities and CO linewidths than the M~33 clouds.  
The specific angular momentum of the M~51 clouds appears to increase with mass only as $j(M) \propto M^{0.4}$ whereas the slope is steeper for M~33 $j(M) \propto M^{0.68}$ (selecting the high S/N clouds).  The uncertainties on the slopes of the fits of $j(M)$ are 0.09 and 0.08 such that the difference is only $1.5\sigma$, which we do not consider significant.  For a cloud mass of $10^6$ M$_\odot$, the fits yield $50 \kms$ pc for M~51 and $45  \kms$ pc for M~33. 

The specific angular momentum in the M51 clouds varies sublinearly (0.4) with cloud mass but the distribution shows a lot of scatter.  In the M~33 clouds, the variation, again with a scatter of a factor of several, is also slightly sublinear (0.68).
This is in excellent agreement with Figure 5 of \citet{Phillips99} for Galactic clouds (slope of 0.72) despite the absence of overlap in cloud masses (the Phillips clouds are smaller and less massive).  It is also worth noting that the extragalactic clouds (M51 and M33, here and in B18) are an extension of the values of specific angular momentum found by \citet{Phillips99}.  The trends found by \citet{Phillips99} and by B18 for M~33 are shown as thick solid or dashed green lines.  Extending the \citet{Phillips99} trend to $10^6$\,M$_\odot$ yields $j=79 \kms$ pc, higher but well within the scatter of the extragalactic values.

The bottom plot of Fig. \ref{angmom2} shows how the specific angular momentum varies with cloud size.  While the slope of $j(M)$ was slightly steeper for Galactic clouds \citep{Phillips99} than M~51 (although similar to M~33), the reverse is true for $j(R)$ where \citet{Phillips99} found 1.3 but the slope is 1.6 for M~51 and $\sim 1.8$ for M~33.  Although the slopes of $j(R)$ are not significantly different between M~33 and M~51, the angular momentum for a given cloud size is higher (significance $>3 \sigma$) in the M~51 clouds.
It is more difficult to compare with the Galactic clouds because they are smaller and the  \citep{Phillips99} sample is quite inhomogeneous so the definition of "radius" may not correspond exactly to what CPROPS measures.  In a situation where the galactic differential rotation contributes at the large scales, 
it is not surprising that we find a steeper $j(R)$ compared to the smaller scales in \citet{Phillips99}. 
Alternatively, the steepening at large (full GMC) scales suggests that galactic scale dynamics has an influence.

All of these features argue in favor of rotation.
However, 
there is no obviously preferred orientation for the rotation axis with respect to the cloud major axis determined by CPROPS.  In fact, the difference between the rotation and morphological axes appears to be distributed randomly.

It is quite remarkable that over several orders of magnitude in mass and a factor $\sim 100$ in size, similar specific angular momentum trends are observed.  
The similar scaling relationships likely point to a driving physical origin.  Any systematic cloud rotation should be driven by the galaxy and a signature of cloud rotation would be the alignment of the cloud angular momenta with the overall angular momentum of the galaxy (i.e., prograde cloud rotation). 
A turbulent velocity field of the molecular medium with a steep power-law power spectrum could lead to the largest-scale velocity modes having the most power, which could manifest itself as a gradient \citep{Burkert2000}, although not necessarily prograde.  The two explanations could be linked as galactic rotation could drive rotational motions in the molecular medium that exhibit common properties across a range of scales from a turbulent velocity cascade.


From Table 1 and Fig. \ref{angmom2} it is apparent that the $j(M)$ relations for M~33 and M~51 are similar and both below the  \citet{Phillips99} values for the Galaxy.  However, one can also see that the $j(R)$ is higher for M~51, presumably due to the higher gradient in gravitational potential.

\begin{table}
\begin{center}
\begin{tabular}{llllr}
property  & M~33 & rms & M~51 & rms \\
\hline
$j(M)_{all} =$ & $48.9 \ (M/10^6\msol)^{0.51}$ & 3.0 &  $53.4 \ (M/10^6\msol)^{0.39}$ & 2.4   \\
$j(M)_{strong} =$ &  $42.6 \ (M/10^6\msol)^{0.68}$ & 3.0 &   $47.3 \ (M/10^6\msol)^{0.42}$ & 2.4    \\
$j(R)_{all} =$ &  $42.0 \ (R/60\pc)^{1.82}$ & 2.1 &  $86.7 \ (R/60\pc)^{1.55}$ & 1.9    \\
$j(R)_{strong} =$ & $37.1\ (R/60\pc)^{1.90}$ & 2.1 &   $79.3 \ (R/60\pc)^{1.63}$ & 1.9  \\
\end{tabular}
\caption[]{Scaling relations for the specific angular momentum $j$ as a function of cloud mass and size.  The relations found by \citet{Phillips99} are $j(M) = 79.4 \ (M/10^6\msol)^{0.72}$ and $j(R) = 49.9 \ (R/60\pc)^{1.27}$.  The scatter (rms) is a factor indicating the deviation from the relation given in the table, typically a factor 2 or 3.  The normalizations at $10^6\msol$ and $60\pc$ were chosen as typical of the extragalactic clouds.  
}
\end{center}
\end{table}

\section{Link between velocity gradients and rotation}

\begin{figure*}
	\centering
	\includegraphics[width=\hsize{}]{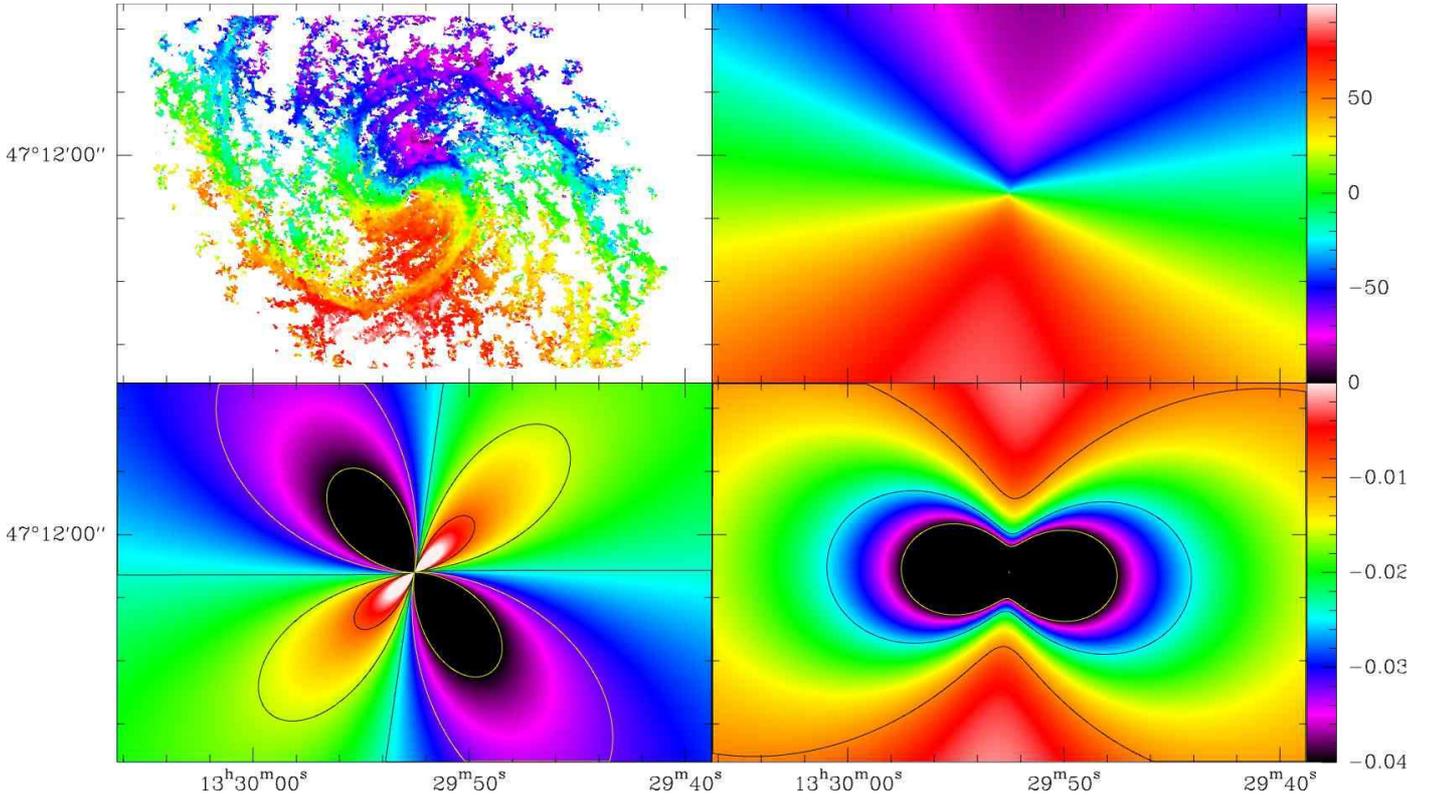}
	\caption{ The first panel shows the PAWS first moment (velocity) from the CO(1--0) observations by \citet{Schinnerer2013} and the upper right panel shows an analytical velocity field designed to approximate the observations but with purely circular rotation and a constant position angle and inclination.  The lower panels show (left) the velocity gradient $\frac{\partial v}{\partial x}$ and (right) $\frac{\partial v}{\partial y}$ due to rotation.  Contours are at -20, -10, 0, 10, and 20 $\mspc$ for $\frac{\partial v}{\partial x}$ and at -10, -25, and -40 (yellow) $\mspc$ for $\frac{\partial v}{\partial y}$.  The color bars show (top) the velocity and (bottom) the velocity gradient for $\frac{\partial v}{\partial y}$.  For $\frac{\partial v}{\partial x}$ (lower left panel), the straight lines are the 0 $\mspc$ contour which defines the regions discussed in Sect. 3.  All velocity gradients are measured in the sky plane (RA-Dec). A more complete explanation of this type of figure can be found in Section 5.1 of B18.}
	\label{dvdxy} 
\end{figure*}

Figure \ref{dvdxy} shows the observed and modeled velocities in the upper panels.  The central part is not well reproduced because there is an apparent change in the position angle whereas the model is axisymmetric.  The model is used to calculate how the observed velocity due to galactic rotation changes when moving along the RA or Dec axes.  In other words, the $\frac{\partial v}{\partial x}$ and $\frac{\partial v}{\partial y}$ coefficients are
calculated for each position in the galaxy due to rotation based on the model.  The $\frac{\partial v}{\partial x}=-\frac{\partial v}{\partial RA}$ and $\frac{\partial v}{\partial y}=\frac{\partial v}{\partial Dec}$ are given in the lower panels.

To further investigate the link between galactic rotation and the cloud velocity gradients, which only cover a very small fraction of the surface of M~51, we calculate the average {\it cloud} $\frac{\partial v}{\partial x}$ where the corresponding galactic coefficient is positive or negative (lower left panel).  The cloud $\frac{\partial v}{\partial x}$ follows the galactic rotation as the average cloud $\frac{\partial v}{\partial x} = 0.008\kms$ pc$^{-1}$ where the galactic $\frac{\partial v}{\partial x}$ is positive and $\frac{\partial v}{\partial x} = -0.0052\kms$ pc$^{-1}$ where  the galactic $\frac{\partial v}{\partial x}$ is negative.  Along the Declination axis, all galactic values are negative because the velocity gradient of M~51 is positive toward the south (see bottom right panel).  Taking regions where $\frac{\partial v}{\partial y}=\frac{\partial v}{\partial Dec} > -0.025\kms$ pc$^{-1}$, the average of the clouds is $\frac{\partial v}{\partial y} = -0.0137\kms$ pc$^{-1}$.
Where the galactic $\frac{\partial v}{\partial y} < -0.025\kms$ pc$^{-1}$, the clouds have on average $\frac{\partial v}{\partial y} = -0.0252\kms$ pc$^{-1}$.  These values are for the S/N$\ge 10$ sample but similar values are found for lower S/N samples.  Thus, in all cases the average rotation is prograde but lower than the galactic gradient or shear.  Our measurements are in a frame not rotating with the galaxy -- an inertial frame.  When a cloud becomes gravitationally bound, it feels the shear but is not torn apart by it, meaning that the cloud rotation is below that of the shear (as observed here and in M~33).  The average galactic shear at the positions of the clouds is 
$\frac{\partial v}{\partial x}=-0.0198\kms$ pc$^{-1}$ for the region with $\frac{\partial v}{\partial x} \le 0$, 
$\frac{\partial v}{\partial x}=0.0132\kms$ pc$^{-1}$ for the region with $\frac{\partial v}{\partial x} > 0$, 
$\frac{\partial v}{\partial y} = -0.059\kms$ pc$^{-1}$ for the region with $\frac{\partial v}{\partial y} \le -0.025$, and 
$\frac{\partial v}{\partial y} = -0.015\kms$ pc$^{-1}$ for the region with $\frac{\partial v}{\partial y} > -0.025$.
In all cases, this is higher (in absolute value) than the cloud gradients calculated in the same way ($-0.0052$, $0.0080$, $-0.0252$, and $-0.0137\kms$ pc$^{-1}$, respectively).  Thus, the cloud velocity gradients are on average slightly less than half what is provided by shear, showing (if need be) the effect of the cloud self-gravity. 

Throughout the region where these clouds are found, the rotation curve is rising \citep[see Fig. 10 of][]{Meidt2013}.  \citet{Mestel66} showed that throughout the rising part of a rotation curve, conservation of angular momentum implies that the rotation will be prograde.  Clearly this does not include cloud-cloud collisions or winds or other pressure of any origin (i.e., the other possible sources of velocity gradients) and in fact these effects may explain why nearly 1/3 of the strong clouds (S/N$\ga 10$) show retrograde gradients.  We have shown that the sign and amplitude of the velocity gradients follow the sign and amplitude of the shear, as expected from gravitational contraction in a rotating disk.
M~33 also has a rising rotation curve and all the same features were seen by B18, so the present results suggest that these are general features, at least for the rising parts of rotation curves.  In the case of M~51, the strong local spiral arm potential may affect the angular momenta of some clouds.

\section{Link between velocity gradients and spiral arms}

A rather simple test was applied to see whether the spiral potential modified the pro/retrograde orientation of the cloud velocity gradients.  \citet[][Fig. 1]{Colombo2014a} provide a mask defining different environments.  Since the rotation curve is rising everywhere, only the non-axisymmetric potential of the spiral arms could cause retrograde rotation in clouds formed through gravitational condensation.  Comparing the arm and "not-arm" regions, we find that retrograde rotation is indeed considerably higher in the arm regions.  For the S/N$\ge7$ sample, retrograde rotation represents 40\% in the arms vs 30\% in the not-arm regions.  For the S/N$\ge10$ sample, retrograde rotation represents 36\% in the arms but only 27\% in the not-arm regions.  The average velocity gradient amplitudes are similar at about $0.05\kms$ pc$^{-1}$ for the 4 categories (arm-retro, arm-pro, notarm-retro, and notarm-pro). Thus, while a majority of the clouds show prograde rotation, the fraction of retrograde rotators is higher where there is a strong non-axisymmetric component to the gravitational potential.  The results of the comparison of the regions of M~51 are summarized in Table 2.

\begin{figure*}
	\centering
	\includegraphics[width=\hsize{}]{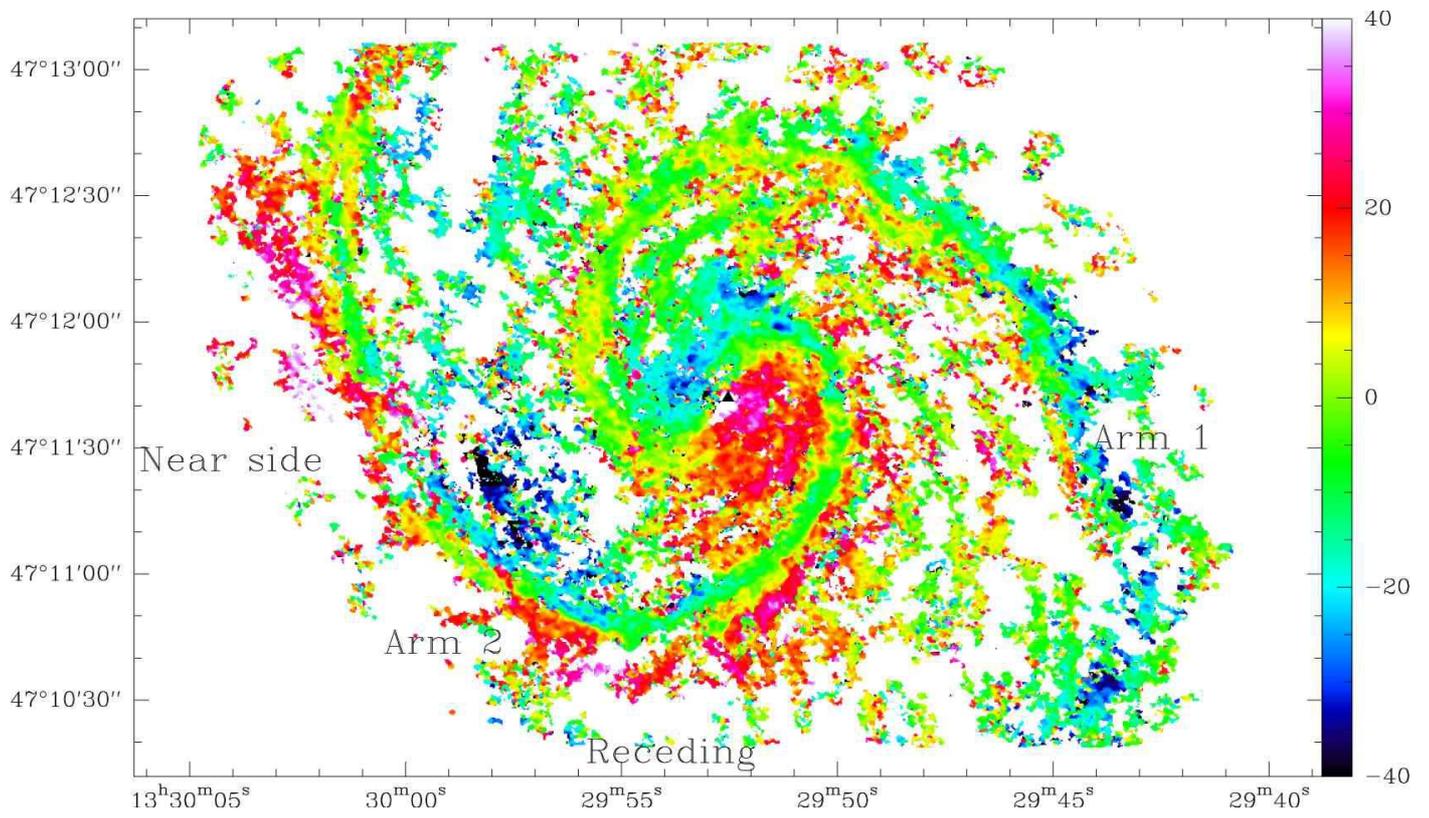}
	\caption{ Residual velocity field as in \citet{Colombo2014b}, with annotations to clarify text.
	Note how each arm has a red-shifted and a blue-shifted side; this is used to separate the two sides of the arms. }
	\label{paws_res} 
\end{figure*}

The corotation radius of M~51 is at about 100$"$ from the center \citep{Querejeta2016}, so the clouds discussed here
are within corotation.  Being within corotation means that the stars and clouds go around the center more quickly than the arm pattern.  Clouds enter the concave side of the arms and exit through the convex leading edge.  They are accelerated when entering and slowed down upon leaving the arms.  This explains the coherently red and blue shifted regions around the arms seen in Fig.  \ref{paws_res}.  

We took the residual velocity map from Fig. 1 (top) of \citet{Colombo2014b} in which streaming motions are clearly visible and tested for the number of pro and retrograde rotators in the leading and trailing parts of the arms.
In the western spiral arm (Arm 1 in Fig.  \ref{paws_res}), mostly on the far side of M~51, the leading edge is blue-shifted and the trailing (concave) edge is red-shifted compared to the large-scale rotation curve.  In the eastern spiral arm (Arm 2), the situation is reversed.  This can be seen in Fig. \ref{paws_res}.
We identified three regions: arm leading, arm trailing, and "not-arm" as a control sample.
The majority of the clouds in the leading parts of the spiral arms are retrograde rotators!  This is true for both the S/N$\ge10$ and S/N$\ge7$ samples.  Less than 25\% of the clouds on the trailing edges are retrograde rotators.
About 30\% of the not-arm clouds are retrograde rotators and these are distributed equally among regions where the velocity residual is red or blue-shifted.

\begin{figure}
	\centering
	\includegraphics[width=\hsize{}]{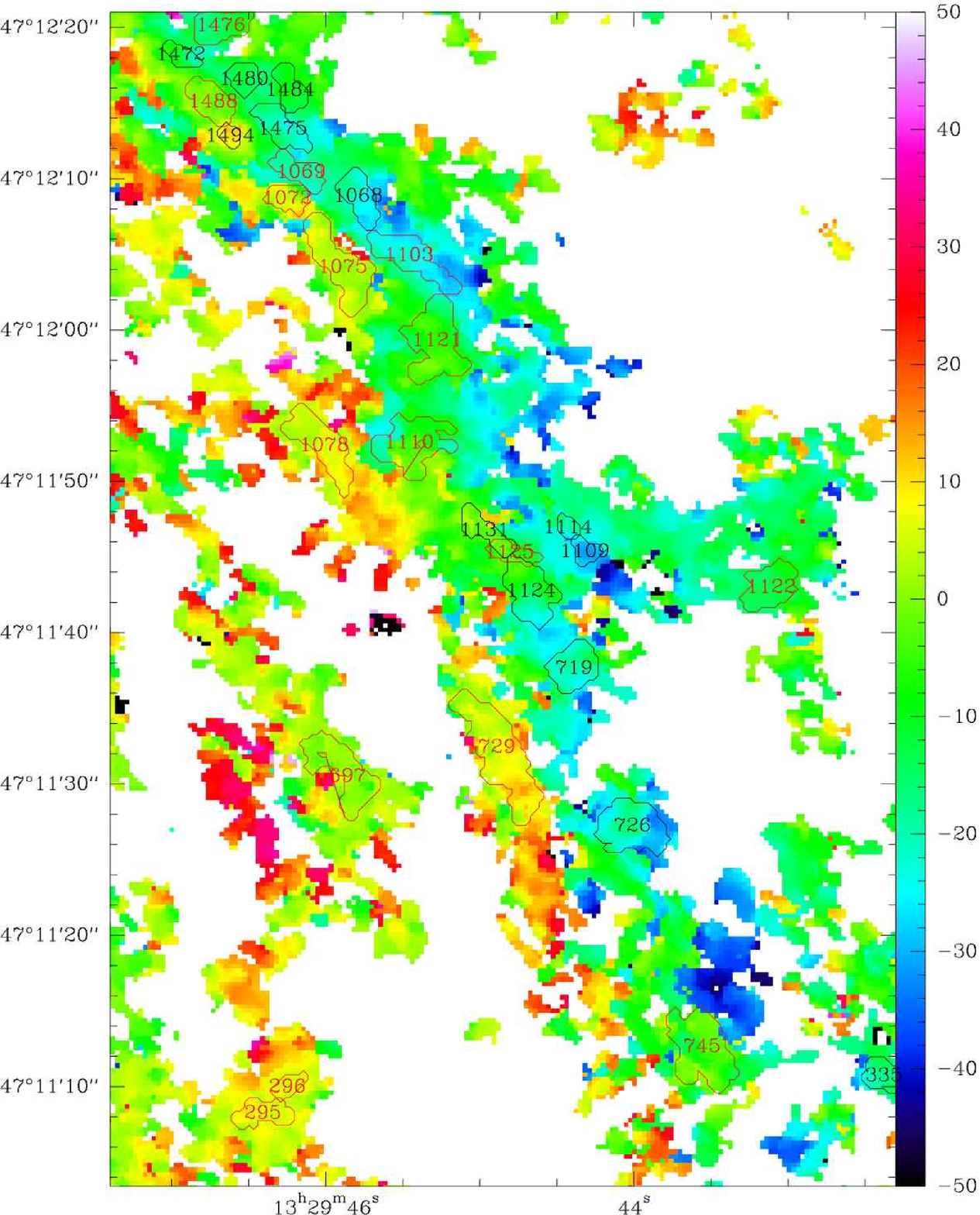}
	\caption{ Residual velocity field as before but only for Arm 1.  Printed in red and black are respectively the prograde and retrograde rotators from the S/N$\ge 10$ sample.  The contours give the cloud sizes and shapes.  As can be seen, most clouds enter the arm (redshifted side for Arm 1) as prograde rotators and come out as retrograde rotators.  The same is seen in Arm 2 and in the lower S/N sample (see last part of Table 2) but there are too many clouds for a simple visualization. }
	\label{paws_res2} 
\end{figure}

Figure \ref{paws_res2} shows this more graphically by focussing on Arm 1.  The clouds, whose positions are indicated by the numbers \citep{Colombo2014a}, enter the arms as prograde rotators (red numbers) but exit as retrograde (black numbers) rotators!  Presumably this is due to the non-axisymmetric gravitational potential of the spiral arm or to cloud destruction-formation mechanisms within the arm.  The stellar arm-interarm contrast is about 3 in M~51 \citep{Elmegreen2011,Rix93} as compared to probably 1.6 in M~33 \citep{Elmegreen2011}, so it is not surprising to see spiral-arm related features in M~51 while not in M~33.  \citet{Meidt2018} show how GMCs can have internally driven motions (turbulence) in the interarm region but their kinematics could be driven (partially) by the large-scale potential upon arrival in the arms. 
This systematic behavior is yet further evidence that the velocity gradients are not due to random effects.

\begin{table}
\begin{center}
\begin{tabular}{lllr}
property  & S/N$=7$& S/N$=10$ & comments \\
\hline
\multicolumn{4}{c}{Whole Galaxy} \\
\hline
prograde & 435 & 257 &     \\
retrograde & 247 & 119 &     \\
\hline
\multicolumn{4}{c}{Regions of Fig.~\ref{dvdxy}} \\
\hline
$\frac{\partial v}{\partial x} > 0$ & 0.0084 &0.0080 & ave cloud grad $\kms$ pc$^{-1}$ \\ 
$\frac{\partial v}{\partial x} < 0$ & -0.0035 & -0.0052 & ave cloud grad $\kms$ pc$^{-1}$ \\ 
$\frac{\partial v}{\partial y} > -0.025$ & -0.0096 & -0.0137 & ave cloud grad $\kms$ pc$^{-1}$\\ 
$\frac{\partial v}{\partial y} < -0.025$ & -0.0217 & -0.0252 & ave cloud grad $\kms$ pc$^{-1}$ \\ 
\hline
\multicolumn{4}{c}{Effect of spiral arms} \\
\hline
arm & 143 & 66 & retrograde\\
arm & 201 & 116 & prograde \\
not arm & 104 & 53 & retrograde \\
not arm & 234 & 141 & prograde \\
\hline
\multicolumn{4}{c}{Effect of location within spiral arms} \\
\hline
leading & 98 & 42& retrograde \\
leading & 69 & 37 & prograde \\
inner & 45 & 24& retrograde \\
inner & 132 & 79 & prograde \\
not arm & 104 & 53 & retrograde \\
not arm & 234 &141 & prograde \\
\end{tabular}
\caption[]{Comparison of rotation in regions of M51, showing the results for both samples. 
Top part shows that rotation is generally prograde.  The second part shows that cloud rotation follows
galactic shear.  The third part shows that the fraction of retrograde rotators is higher in the spiral arms.
The last part shows that the retrograde rotators are mostly found in and actually {\it dominate} the leading edge of the arms. The fraction of pro- and retrograde rotators does not vary between the red and blue shifted parts of the "not arm" regions.}
\end{center}
\end{table}

\section{Conclusions}

Cloud velocity gradients are seen within the PAWS sample of clouds.
Dominantly prograde velocity gradients are observed in the \citet{Colombo2014a} cloud sample.  This in itself, coupled with the similar results found by \citet{Braine2018} for M~33, is evidence that velocity gradients are chiefly due to cloud rotation.  30\% of the clouds show retrograde rotation.  As in M~33, the rotational energy is not sufficient to support the cloud against gravity, at least at the GMC-scale observed here.  

The gradients follow -- but with lower amplitude -- the shear in the disk, which is expected for gravitational contraction within an axisymmetric differentially rotating disk.  Strictly prograde rotation would be expected in M~51 and M~33 for perfect measurements with only gravity in an axisymmetric potential.  Thus, the presence in both galaxies of retrograde gradients indicates that other processes also play a role
in creating the observed gradients, in addition to measurement uncertainties.  The non-axisymmetric potential of the spiral arm regions is one of the causes of retrograde rotation.  Retrograde rotation is more common along the spiral arms than in the disk of M~51 as a whole.
In particular, the leading (i.e., outer) edge of the spiral arms actually shows a majority of retrograde rotators, such that clouds apparently as prograde rotators and leave the arms as dominantly retrograde rotators. 

\bibliographystyle{aa}
\bibliography{jb}

\begin{thebibliography}{24}
\expandafter\ifx\csname natexlab\endcsname\relax\def\natexlab#1{#1}\fi

\bibitem[{{Bally}(1989)}]{Bally89}
{Bally}, J. 1989, in European Southern Observatory Conference and Workshop
  Proceedings, Vol.~33, European Southern Observatory Conference and Workshop
  Proceedings, ed. B.~{Reipurth}, 1--32

\bibitem[{{Blitz}(1993)}]{Blitz93}
{Blitz}, L. 1993, in Protostars and Planets III, ed. E.~H. {Levy} \& J.~I.
  {Lunine}, 125--161

\bibitem[{{Braine} {et~al.}(2018){Braine}, {Rosolowsky}, {Gratier}, {Corbelli},
  \& {Schuster}}]{Braine2018}
{Braine}, J., {Rosolowsky}, E., {Gratier}, P., {Corbelli}, E., \& {Schuster},
  K.-F. 2018, A\&A, 612, A51

\bibitem[{{Burkert} \& {Bodenheimer}(2000)}]{Burkert2000}
{Burkert}, A. \& {Bodenheimer}, P. 2000, ApJ, 543, 822

\bibitem[{{Colombo} {et~al.}(2014{\natexlab{a}}){Colombo}, {Hughes},
  {Schinnerer}, {Meidt}, {Leroy}, {Pety}, {Dobbs}, {Garc{\'{\i}}a-Burillo},
  {Dumas}, {Thompson}, {Schuster}, \& {Kramer}}]{Colombo2014a}
{Colombo}, D., {Hughes}, A., {Schinnerer}, E., {et~al.} 2014{\natexlab{a}},
  ApJ, 784, 3

\bibitem[{{Colombo} {et~al.}(2014{\natexlab{b}}){Colombo}, {Meidt},
  {Schinnerer}, {Garc{\'{\i}}a-Burillo}, {Hughes}, {Pety}, {Leroy}, {Dobbs},
  {Dumas}, {Thompson}, {Schuster}, \& {Kramer}}]{Colombo2014b}
{Colombo}, D., {Meidt}, S.~E., {Schinnerer}, E., {et~al.} 2014{\natexlab{b}},
  ApJ, 784, 4

\bibitem[{{Corbelli} {et~al.}(2017){Corbelli}, {Braine}, {Bandiera},
  {Brouillet}, {Combes}, {Druard}, {Gratier}, {Mata}, {Schuster}, {Xilouris},
  \& {Palla}}]{Corbelli17}
{Corbelli}, E., {Braine}, J., {Bandiera}, R., {et~al.} 2017, A\&A, 601, A146

\bibitem[{{Dobbs}(2008)}]{Dobbs08}
{Dobbs}, C.~L. 2008, MNRAS, 391, 844

\bibitem[{{Druard} {et~al.}(2014){Druard}, {Braine}, {Schuster}, {Schneider},
  {Gratier}, {Bontemps}, {Boquien}, {Combes}, {Corbelli}, {Henkel}, {Herpin},
  {Kramer}, {van der Tak}, \& {van der Werf}}]{Druard14}
{Druard}, C., {Braine}, J., {Schuster}, K.~F., {et~al.} 2014, A\&A, 567, A118

\bibitem[{{Elmegreen} {et~al.}(2011){Elmegreen}, {Elmegreen}, {Yau},
  {Athanassoula}, {Bosma}, {Buta}, {Helou}, {Ho}, {Gadotti}, {Knapen},
  {Laurikainen}, {Madore}, {Masters}, {Meidt}, {Men{\'e}ndez-Delmestre},
  {Regan}, {Salo}, {Sheth}, {Zaritsky}, {Aravena}, {Skibba}, {Hinz}, {Laine},
  {Gil de Paz}, {Mu{\~n}oz-Mateos}, {Seibert}, {Mizusawa}, {Kim}, \& {Erroz
  Ferrer}}]{Elmegreen2011}
{Elmegreen}, D.~M., {Elmegreen}, B.~G., {Yau}, A., {et~al.} 2011, ApJ, 737, 32

\bibitem[{{Gratier} {et~al.}(2012){Gratier}, {Braine}, {Rodriguez-Fernandez},
  {Schuster}, {Kramer}, {Corbelli}, {Combes}, {Brouillet}, {van der Werf}, \&
  {R{\"o}llig}}]{Gratier12}
{Gratier}, P., {Braine}, J., {Rodriguez-Fernandez}, N.~J., {et~al.} 2012, A\&A,
  542, 108

\bibitem[{{Gratier} {et~al.}(2010){Gratier}, {Braine}, {Rodriguez-Fernandez},
  {Schuster}, {Kramer}, {Xilouris}, {Tabatabaei}, {Henkel}, {Corbelli},
  {Israel}, {van der Werf}, {Calzetti}, {Garcia-Burillo}, {Sievers}, {Combes},
  {Wiklind}, {Brouillet}, {Herpin}, {Bontemps}, {Aalto}, {Koribalski}, {van der
  Tak}, {Wiedner}, {Roellig}, \& {Mookerjea}}]{Gratier10}
{Gratier}, P., {Braine}, J., {Rodriguez-Fernandez}, N.~J., {et~al.} 2010, A\&A,
  522, 3

\bibitem[{{Hu} {et~al.}(2013){Hu}, {Shao}, \& {Peng}}]{Hu2013}
{Hu}, T., {Shao}, Z., \& {Peng}, Q. 2013, ApJL, 762, L27

\bibitem[{{Imara} {et~al.}(2011){Imara}, {Bigiel}, \& {Blitz}}]{Imara11b}
{Imara}, N., {Bigiel}, F., \& {Blitz}, L. 2011, ApJ, 732, 79

\bibitem[{{Kutner} {et~al.}(1977){Kutner}, {Tucker}, {Chin}, \&
  {Thaddeus}}]{Kutner77}
{Kutner}, M.~L., {Tucker}, K.~D., {Chin}, G., \& {Thaddeus}, P. 1977, ApJ, 215,
  521

\bibitem[{{Meidt} {et~al.}(2018){Meidt}, {Leroy}, {Rosolowsky}, {Kruijssen},
  {Schinnerer}, {Schruba}, {Pety}, {Blanc}, {Bigiel}, {Chevance}, {Hughes},
  {Querejeta}, \& {Usero}}]{Meidt2018}
{Meidt}, S.~E., {Leroy}, A.~K., {Rosolowsky}, E., {et~al.} 2018, ApJ, 854, 100

\bibitem[{{Meidt} {et~al.}(2013){Meidt}, {Schinnerer}, {Garc{\'{\i}}a-Burillo},
  {Hughes}, {Colombo}, {Pety}, {Dobbs}, {Schuster}, {Kramer}, {Leroy}, {Dumas},
  \& {Thompson}}]{Meidt2013}
{Meidt}, S.~E., {Schinnerer}, E., {Garc{\'{\i}}a-Burillo}, S., {et~al.} 2013,
  ApJ, 779, 45

\bibitem[{{Mestel}(1966)}]{Mestel66}
{Mestel}, L. 1966, MNRAS, 131, 307

\bibitem[{{Phillips}(1999)}]{Phillips99}
{Phillips}, J.~P. 1999, A\&AS, 134, 241

\bibitem[{{Querejeta} {et~al.}(2016){Querejeta}, {Meidt}, {Schinnerer},
  {Garc{\'{\i}}a-Burillo}, {Dobbs}, {Colombo}, {Dumas}, {Hughes}, {Kramer},
  {Leroy}, {Pety}, {Schuster}, \& {Thompson}}]{Querejeta2016}
{Querejeta}, M., {Meidt}, S.~E., {Schinnerer}, E., {et~al.} 2016, A\&A, 588,
  A33

\bibitem[{{Rix} \& {Rieke}(1993)}]{Rix93}
{Rix}, H.-W. \& {Rieke}, M.~J. 1993, ApJ, 418, 123

\bibitem[{{Rosolowsky} {et~al.}(2003){Rosolowsky}, {Engargiola}, {Plambeck}, \&
  {Blitz}}]{Rosolowsky03}
{Rosolowsky}, E., {Engargiola}, G., {Plambeck}, R., \& {Blitz}, L. 2003, ApJ,
  599, 258

\bibitem[{{Rosolowsky} \& {Leroy}(2006)}]{Rosolowsky06}
{Rosolowsky}, E. \& {Leroy}, A. 2006, PASP, 118, 590

\bibitem[{{Schinnerer} {et~al.}(2013){Schinnerer}, {Meidt}, {Pety}, {Hughes},
  {Colombo}, {Garc{\'{\i}}a-Burillo}, {Schuster}, {Dumas}, {Dobbs}, {Leroy},
  {Kramer}, {Thompson}, \& {Regan}}]{Schinnerer2013}
{Schinnerer}, E., {Meidt}, S.~E., {Pety}, J., {et~al.} 2013, ApJ, 779, 42

\end{thebibliography}

\end{document}